\documentclass[%
reprint,
amsmath,amssymb,
prd,
nofootinbib
]{revtex4-2}
\usepackage{xspace}
\usepackage{lipsum}  
\usepackage{amsfonts}
\usepackage{epsfig}
\usepackage{color}
\usepackage{soul}
\usepackage{amstext}
\usepackage{amssymb}
\usepackage{verbatim}
\usepackage{graphicx}
\usepackage{mathrsfs}
\usepackage{wrapfig}
\usepackage[T1]{fontenc}
\usepackage[percent]{overpic}
\usepackage{xcolor}
\usepackage[]{hyperref}
\usepackage{booktabs}
\usepackage{lineno}
\usepackage{orcidlink}
\hypersetup{
	colorlinks,
	linkcolor={blue},
	linktocpage=true,
	citecolor={red},
	urlcolor={olive}
}

\makeatletter
\renewcommand\subsubsection{\@startsection{subsubsection}{3}{\z@}%
                                     {-3.25ex\@plus -1ex \@minus -.2ex}%
                                     {-1.5ex \@plus -.2ex}
                                     {\normalfont\normalsize\bfseries}}
\makeatother

\def\epem       {\ensuremath{e^+e^-}\xspace}
\def\ffbar      {\ensuremath{f\overline f}\xspace}
\def\ra         {\ensuremath{\rightarrow}\xspace}

\def\stw       {\ensuremath{\sin^2\theta_W}\xspace}
\def\alr       {\ensuremath{A_{\textrm{LR}}}\xspace}
\def\salr       {\ensuremath{A_{\textrm{LR}\Sigma} }\xspace}

\usepackage{relsize}
\def\babar{\mbox{\slshape B\kern-0.1em{\smaller A}\kern-0.1em
    B\kern-0.1em{\smaller A\kern-0.2em R}}\xspace}

\begin{document}

{\pagestyle{empty}
\title{Left-right asymmetry calculation comparisons and projected sensitivity to the weak mixing angle in polarized Bhabha scattering at 10.58~GeV}

\author{C.~Miller\orcidlink{0000-0003-2631-1790}, J.M.~Roney\orcidlink{0000-0001-7802-4617}}
\affiliation{University of Victoria, Canada}
\begin{abstract}
Consideration is being given to upgrading the SuperKEKB electron-positron collider with the introduction of electron-beam polarization. This would enable a unique precision electroweak physics program that opens new ways to search for physics beyond the Standard Model.
The upgrade would enable Belle II to make a number of high precision measurements, one of which is the left-right cross-section asymmetry in the $\epem\rightarrow\epem$ Bhabha scattering process.  The expected level of precision in such a measurement will require the theoretical values of the asymmetry to be calculated at least to the next-to-leading order (NLO) level, and the implementation of simulation event generators with a similar level of precision. In this study, we compare the calculations of the ReneSANCe Monte Carlo generator with those of an independent NLO calculation to determine the level of agreement in this process. An average difference of 0.3\% between the calculations is found. Using the published Belle II efficiency for selecting Bhabha events and assuming a  40~ab$^{-1}$ dataset having 70\% polarization, the projected uncertainty on the weak mixing angle, \stw, is calculated using ReneSANCe to be $\pm0.00032$ or better. 
Combining this with left-right asymmetry measurements from muons and taus under the assumption of lepton universality yields a projected overall uncertainty of $\pm0.00019$ on \stw with SuperKEKB upgraded to have polarized electron beams.

\end{abstract}		

\maketitle

\vfill

}
	

\setcounter{footnote}{0}

\section{Introduction}
A proposed upgrade to the SuperKEKB electron-positron collider would introduce polarization to the electron beam~\cite{polarizationWhitePaper}. 
The upgrade, referred to as `Chiral Belle', is designed to introduce polarization to the collider without compromising SuperKEKB's ability to reach its design luminosity.
The addition of a polarized beam opens a new and unique precision program that enables Belle~II to determine the weak mixing angle, \stw, at a 10.58~GeV center-of-mass energy by measuring the left-right cross-section asymmetry, $A_{\textrm{LR}}$, separately for electrons, muons, taus, b-quarks, and c-quarks. Included in its broader program, Chiral Belle will also measure the $\tau$-lepton anomalous magnetic moment at a level of precision two orders of magnitude higher than has been achieved to date. The Chiral Belle measurement of $A_{\textrm{LR}}$ for the $\epem\ra\epem(\gamma)$ process (Bhabha scattering) will complement the proposed measurement of \stw by the MOLLER experiment~\cite{MOLLER:2014iki}, which is projected to determine \stw from electron-electron scattering at the 100~MeV energy scale with an uncertainty of $\pm0.00028$~\cite{Demiroglu:2024wys}. Chiral Belle and MOLLER will probe the running of \stw via $Z^0$-electron couplings at two different energies well below the $Z^0$-pole  and will thereby be sensitive to dark sector or other physics beyond the SM that modifies the running of \stw ~\cite{s2wplot}. The running will also be probed from electron-hadron scattering asymmetries with less precision at the Electron-Ion Collider(EIC)~\cite{Accardi:2012qut}.

By measuring ratios of $A_{\textrm{LR}}$ separately for five different species of fermion, the Chiral Belle program will also provide world-leading measurements of the universality of the $Z^0$-fermion couplings. This paper focuses on predictions of $A_{\textrm{LR}}$ from Bhabha scattering at 10.58~GeV and projections of the uncertainties on \stw associated with $Z^0$-electron couplings that can be obtained with Chiral Belle. 

In order to prepare for and perform these measurements, the Belle II collaboration will require Monte Carlo (MC) event generators, which, at a minimum, fully account for the next-to-leading order (NLO) effects. The ReneSANCe MC generator~\cite{renesance} is one such NLO generator, which also accounts for the initial state polarization states in the $\epem\ra\epem(\gamma)$ process and radiative effects. In presenting a comparison of the ReneSANCe calculations with an independent NLO calculation\cite{ALRee} and making projections for Chiral Belle's expected sensitivity to $A_{\textrm{LR}}$ and \stw, full NNLO calculations of this process are motivated. 

{\bf The Left-Right Electroweak Asymmetry:}
Due to the presence of the weak interaction, the \epem\ra\ffbar process has an asymmetry in the cross sections for left-handed and right-handed beam polarizations. This left-right asymmetry ($A_{LR}$) is defined as~\cite{SuperB,ALRee}:
\begin{equation}
    A_{LR}=\frac{\sigma_L-\sigma_R}{\sigma_L+\sigma_R}
    \label{eqn:alr}
\end{equation}
where $\sigma_{L}(\sigma_{R})$ is the left(right) handed cross-section. The Standard Model (SM) provides predictions of \alr. For example, at Born level in the $s$-channel for the \epem\ra\ffbar  process: 
\begin{equation}
   A_{LR}^{\textrm{meas}} = A_{LR}
   \langle P\rangle
    =\frac{sG_F}{\sqrt{2}\pi\alpha Q_f}g^e_Ag^f_V\langle P\rangle,
    \label{eqn:alr2}
\end{equation}
where $s$ is square of the center-of-mass energy, $G_F$ is the Fermi constant, $\alpha$ is the fine structure constant, $Q_f$ is the electric charge of the final state fermion, $g^e_A$ is the neutral current axial coupling of the initial state electron, $g^f_V$ is the neutral current vector coupling of the final state fermion ($f$):
\begin{align}
    g^f_V&=T_3-2Q_f\sin^2\theta_W\\
    g^e_A&=T_3,
\end{align}
where $T_3$ is the weak isospin ($\pm1/2$) and $Q_f$ is the charge of the fermion. The $\langle P\rangle$ factor in Eqn. \ref{eqn:alr2}  is the average longitudinal polarization of the mediator in the \epem collision, defined as:
\begin{align} \label{eqn:pol_def}
    \langle P\rangle=\frac{R^+L^--L^+R^-}{R^+L^-+L^+R^-},
\end{align}
where $L^{\pm}$ ($R^\pm$) is the fraction of positrons ($+$) or electrons ($-$) in their respective beams that have left-handed (right-handed) spin, so that $L^\pm+R^\pm\equiv1$. Chiral Belle plans to operate with $\langle P\rangle$=70\% and to switch the beam polarization between left and right polarization states during data taking, in order to collect the left and right polarized data sets from which to extract $A_{LR}$. Compton polarimetry is expected to yield real-time measurements of the polarization at the sub-1\%  level every five minutes~\cite{Charlet:2023qvh}. The polarization will also be precisely measured at the interaction point using the tau decay polarimetry technique developed by \babar~\cite{babarpol}.

\begin{figure}[ht]
    \centering
    \begin{minipage}{0.25\textwidth}
        \centering
        \includegraphics[width=\textwidth]{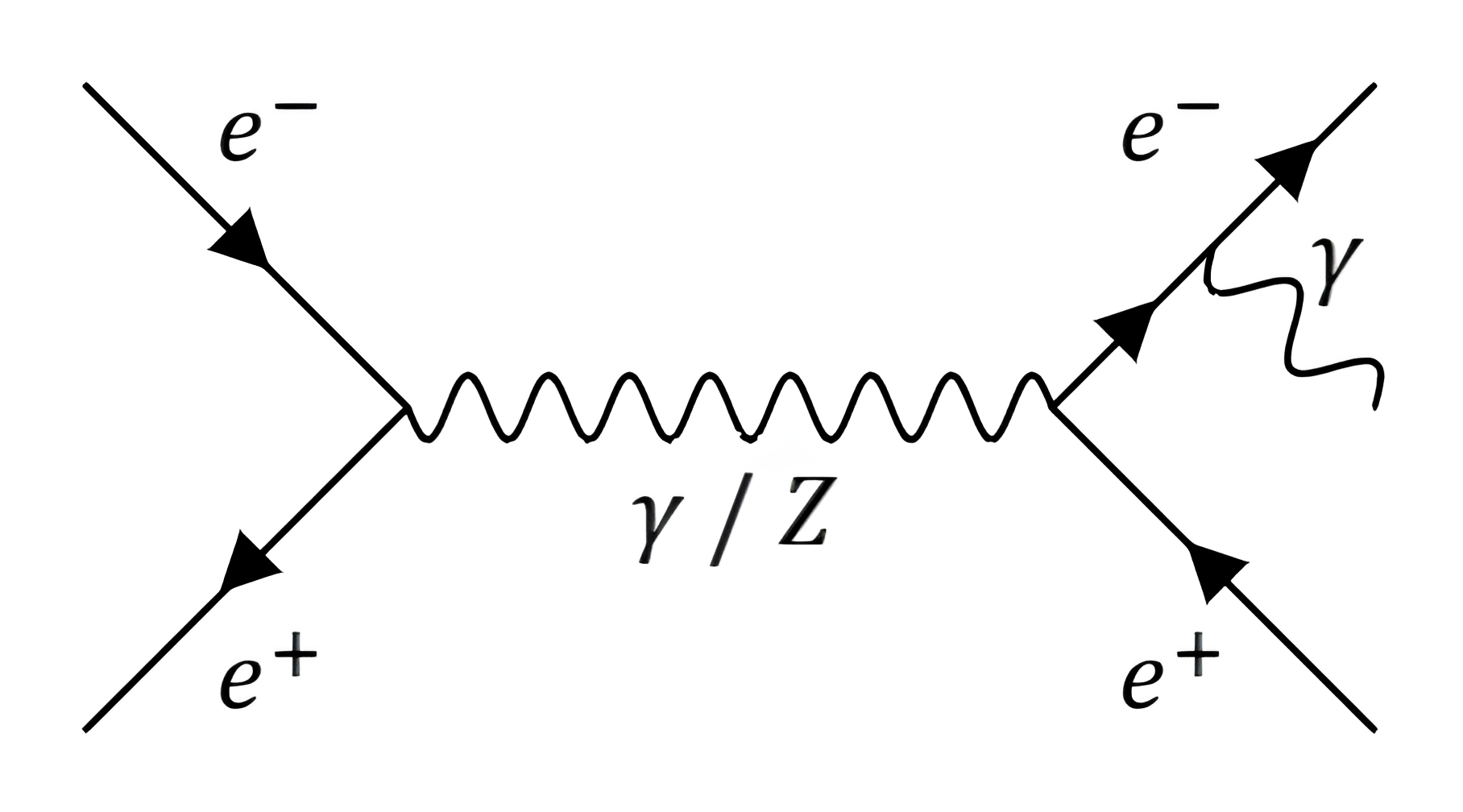}
    \end{minipage}
    \begin{minipage}{0.15\textwidth}
        \centering
        \includegraphics[width=\textwidth]{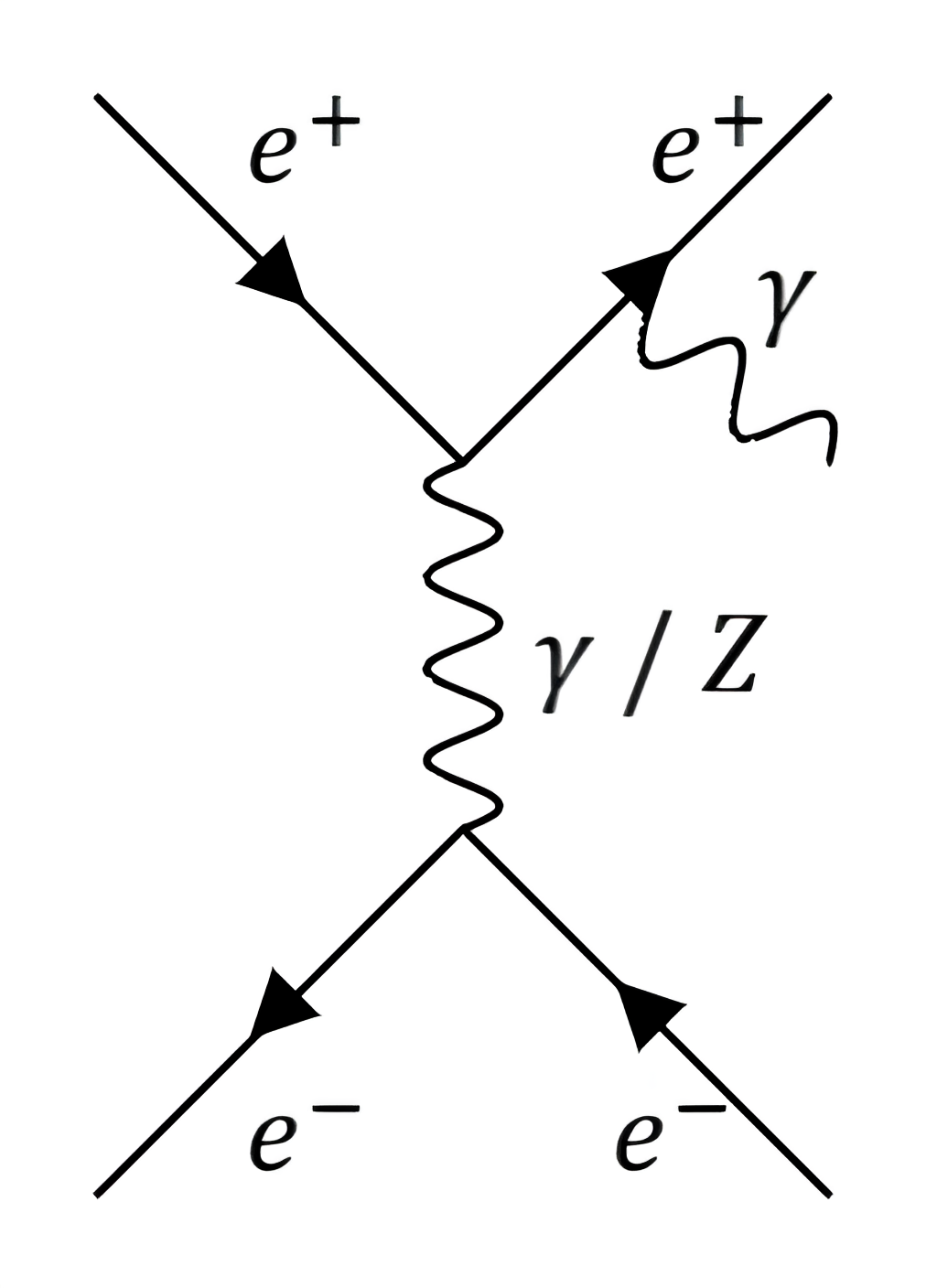}
    \end{minipage}
    \caption{Sample $\epem\ra\epem\gamma$ Feynman diagrams of the s-channel (left) and t-channel (right) processes. Note that the photon can be emitted from any electron or positron.}
    \label{fig:epem_epem_st}
\end{figure}

At a 10.58~GeV center-of-mass energy, both the s-channel and t-channel Bhabha processes, depicted in Figure \ref{fig:epem_epem_st}, contribute significantly to the value of $A_{LR}$.
The Belle II center-of-mass energy of 10.577 GeV is well below the Z-pole which leads to $\gamma$-Z electroweak interference terms dominating the asymmetry. At Born level~\cite{ALRee}:
\begin{equation}
	A^{0}_{LR}=\frac{4v^Z_ea^Z_ess^2_\theta}{m^2_Z}\times\frac{c_\theta(21+c^2_\theta)+12l_\theta}{c_\theta(c_\theta^4+26c^2_\theta-75)-24s^2_\theta l_\theta},
	\label{eqn:alrgg}
\end{equation}
where $\theta$ is the angle between the initial state electron momentum vector and the final state electron momentum vector, $s_\theta$ is $\sin\theta$, $c_\theta$ is $\cos\theta$, and $l_\theta$ is $\ln\frac{1-c_\theta}{1+c_\theta}$.  The $v^Z_e$ and $a^Z_e$ terms are defined as:
\begin{align}
    v_e^Z&\equiv\frac{g^e_V}{2\sin\theta_W\cos\theta_W} \\ 
    a_e^Z&\equiv\frac{g^e_A}{2\sin\theta_W\cos\theta_W}.
\end{align}
In Figures 2 and 3 of Ref.~\cite{ALRee} both \alr and \salr at 10.58~GeV are presented, where \salr is \alr integrated over an angular acceptance:
\begin{equation}
    \salr=\frac{\Sigma_L-\Sigma_R}{\Sigma_L+\Sigma_R}
    \label{eqn:salrdef}
\end{equation}
and
\begin{equation}
    \Sigma_{L(R)}=\int^{\cos a}_{-\cos a}\frac{d\sigma_{L(R)}}{d\cos\theta}\cdot d\cos\theta.
    \label{eqn:salrint}
\end{equation}
In Ref.~\cite{ALRee} \alr and \salr are plotted for values of $\theta$, for the electron, ranging from $5^\circ$ to $85^\circ$ while constraining the positron to $|\cos\theta_{e^+}|\leq\cos20^\circ$. 
\section{R\MakeLowercase{ene}SANC\MakeLowercase{e} MC Generator} 
The ReneSANCe generator was developed to provide simulated events for electron-positron collisions extending to the TeV energy regime at full NLO with radiative corrections~\cite{renesance}. By default the generator accounts for the polarization of the initial state particles and allows them to be set symmetrically to a polarization ranging from 0 to 1. As Chiral Belle will feature a single polarized beam, the ReneSANCe authors added asymmetric beam polarization states as a feature to enable this study.\\
\indent In order to calculate \salr, ReneSANCe can calculate the difference and sum of the left and right-handed cross-sections directly (the numerator and denominator terms seen in Eqn. \ref{eqn:alr}). Using this method the \salr can be calculated for various angular acceptances.

{\bf Generator Settings:} In order to compare the ReneSANCe calculations to those of Ref.~\cite{ALRee}, the following ReneSANCe default SM  parameters were changed to correspond to those in Ref.~\cite{ALRee}: the width of the W boson and the top quark are set to 0, $m_{H}=125$~GeV, $m_Z=91.1876$~GeV, $m_W=80.4628$~GeV, $m_u=69.83$~MeV, $m_d=69.84$~MeV, $m_c=1.2$~GeV, $m_s=0.15$~GeV, $m_t=174$~GeV, and $m_b=4.6$~GeV. The center-of-mass energy is set to 10.577 GeV and the soft photon cutoff, defined as $ome*\sqrt{s}/2$, is set to $ome$=0.002.

\section{\texorpdfstring{\alr}{ALR} Comparisons}
In order to make the comparisons between the ReneSANCe generator output and the calculations from Ref.~\cite{ALRee}, the points plotted in Figure 2 
(\alr  ~{\it vs} $~\theta$) 
and Figure 3 (\salr ~{\it vs} ~$a$, where $a$ is defined in Eqn.~\ref{eqn:salrint}) of Ref.~\cite{ALRee} were extracted graphically, and an uncertainty of 0.2\% assigned from the average value of Table 4 of Ref. ~\cite{ALRee}.

\begin{figure}[ht]
    \centering
	\includegraphics[width=\linewidth]{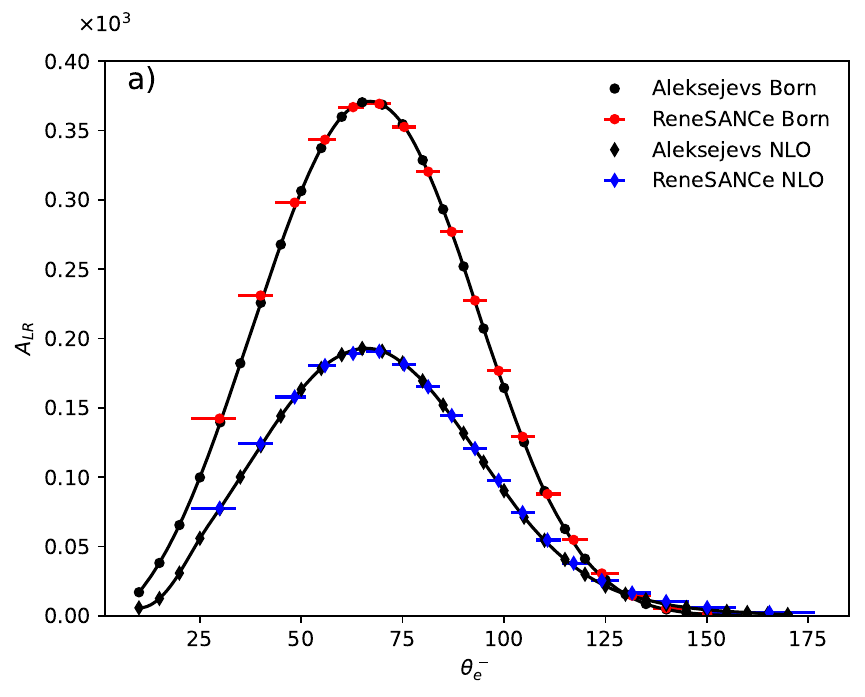}
    \includegraphics[width=\linewidth]{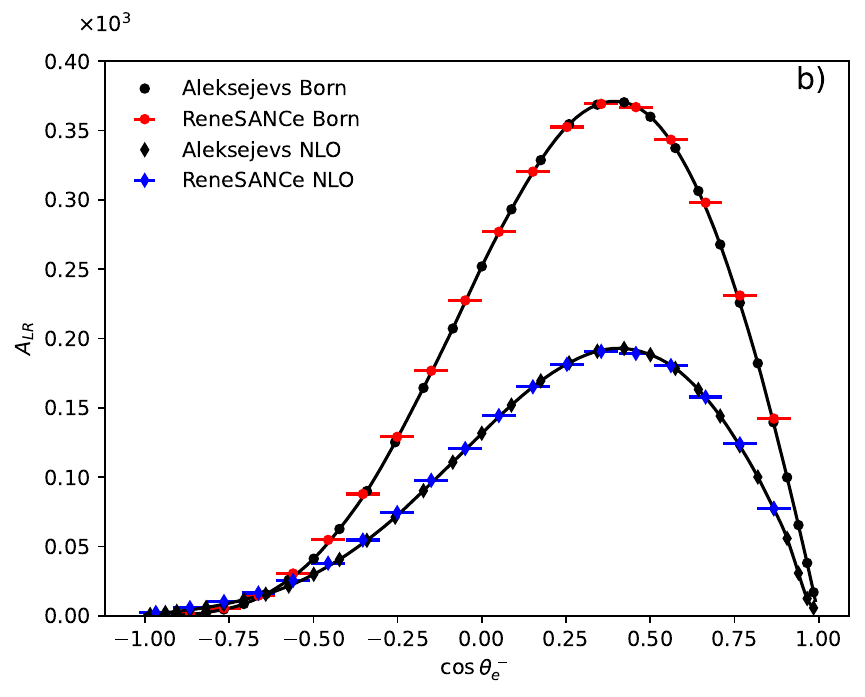}
    \caption{Comparison of the calculations of \alr in Bhabhas from Ref.~\cite{ALRee} and ReneSANCe. A cubic spline is employed to illustrate the trend. The horizontal error bars represent the bin width of $\cos\theta=0.10$. The exponential nature of the Bhabha cross-section results in the central value being biased towards the edge with higher statistics. Sub-figure a) shows the distribution as a function of the final-state electron angle as presented in Fig. 2 of Ref.~\cite{ALRee}, while sub-figure b) shows the same data distributed in $\cos\theta$.}
	\label{fig:alr}
\end{figure}

In ReneSANCe, \alr is calculated by finding \salr in bins of width 0.1 in $\cos\theta$ between $-1<\cos\theta<1$. The central value of the bin reflects the average $\theta$ of events in the bin and the calculated \salr value is the average value of \alr in the bin. The results are shown in Figure \ref{fig:alr}.

\begin{figure}[ht]
    \centering
	\includegraphics[width=\linewidth]{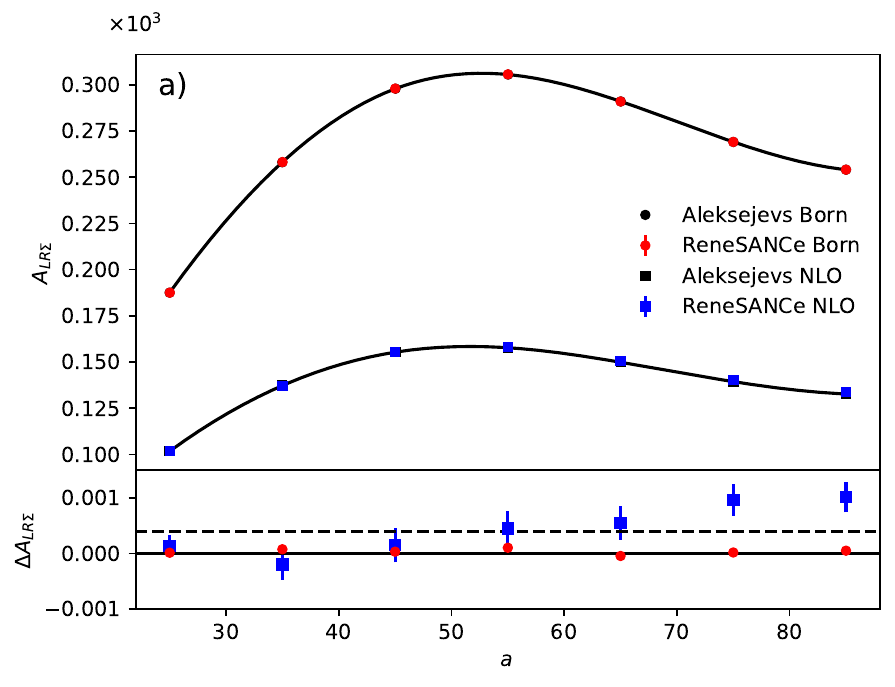}
    \includegraphics[width=\linewidth]{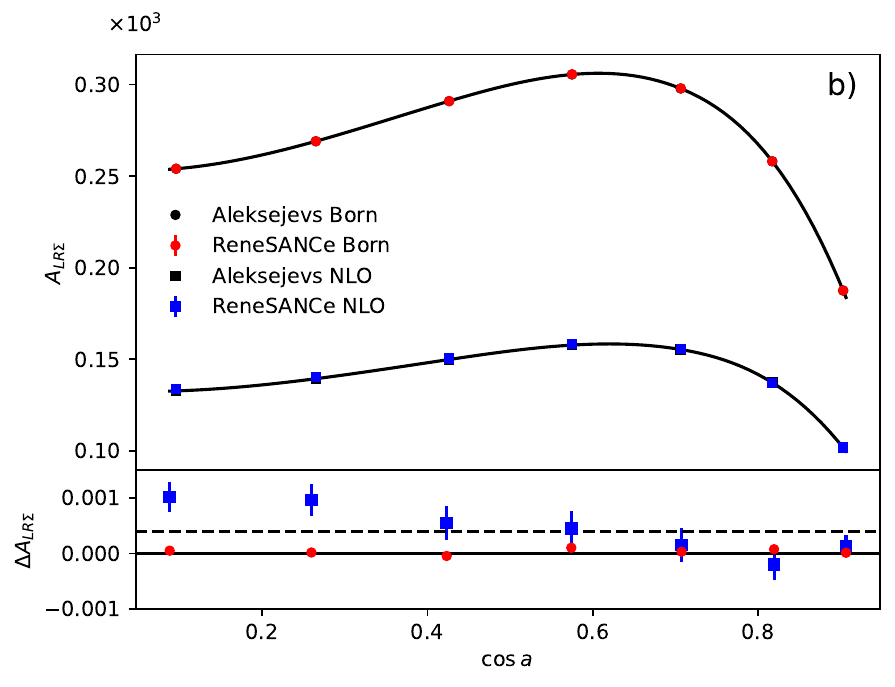}
    \caption{
    Comparison of the calculations of \salr in Bhabhas from Ref.~\cite{ALRee} and ReneSANCe for an angular acceptance integrated between $a$ and $180^\circ-a$. A cubic spline is employed to illustrate the trend. The lower half of each plot shows the absolute difference between the Ref.~\cite{ALRee} and ReneSANCe calculations, with a $\Delta$\salr=4.4$\times10^{-7}$ horizontal line shown as the mean difference. Sub-figure a) shows the distribution as a function of the final-state electron angle as presented in Fig. 2 of Ref.~\cite{ALRee}, while sub-figure b) shows the same data distributed in $\cos\theta$.}
	\label{fig:salr}
\end{figure}

For \salr the symmetric angular acceptance as defined by equations Eqns.~\ref{eqn:salrdef} and \ref{eqn:salrint} is varied from $a=5^\circ$ to $a=85^\circ$ in steps of $10^\circ$. 
The two calculations are overlaid in Fig. \ref{fig:salr}. An average difference of 4.4$\times10^{-7}$, equivalent to a relative difference of 0.3\%, is evident. The observed difference exceeds any uncertainties associated with extracting the points from the plots in \cite{ALRee} and MC statistics from running ReneSANCe. Both Ref.~\cite{ALRee} and ReneSANCe compare their calculations to other sources, with WHIZARD in common between them. These comparisons are carried out for the Born level process and with the addition of radiative effects, and are all in agreement. We therefore assume the difference arises from the calculation of the virtual/internal NLO contributions. However, this difference is also negligible for the purposes on this study.\\
\section{Estimate of NNLO scale in \salr}
The scale of the NNLO contributions is conventionally estimated as $\alpha \times$NLO contribution, i.e. on the order of 1\% of the NLO contributions, which translates into an uncertainty on the asymmetry of  $\delta A_{\textrm{LR}\Sigma}^{\textrm{NNLO}}\sim1.5\times10^{-6}$. 
The NNLO contributions to $\epem\rightarrow\tau^+\tau^-$ cross-section have been calculated and reported in Table 1.~of Ref.~\cite{Kollatzsch:2022bqa}. Their results show the NLO effects act as a $\sim$18\% correction to the Born level cross-sections, while the NNLO effects contribute a $\sim$0.5\% correction to the Born level (or a $\sim$3\% correction to the NLO values). However the lack of the t-channel in $\epem\rightarrow\tau^+\tau^-$ limits the value of using these calculations in providing an estimate for NNLO contributions to Bhabha scattering. NNLO calculations have also been carried out for the MOLLER experiment for polarized e$^-$e$^-$ scattering, which has t and u-channel processes but no s-channel, and the NNLO quadratic and reducible contributions were found to contribute a $\sim$5\% correction to the asymmetry (equivalently a $\sim$12\% correction to the NLO contribution)~\cite{aleks_private,Aleksejevs:2011de, Aleksejevs:2011sq,Aleksejevs:2012zz,Aleksejevs:2015dba,Aleksejevs2016}. The irreducible NNLO contributions are expected to reduce the size of the correction slightly, but it is expected to remain of order 1\%\cite{aleks_private}.  Based on these calculations, the scale of the NNLO effects on the Bhabha scattering \salr are expected to be significant compared to the projected experimental uncertainties from Chiral Belle, discussed in the following section. Therefore a dedicated NNLO calculation of this process is required.

\section{Experimental Expectations}
To estimate the sensitivity Chiral Belle will have to \salr measurements the following assumptions are made. Belle II reports a Bhabha cross-section of 17.4 nb with a detection efficiency for Bhabha events of 0.3593 for an angular acceptance of $|\cos\theta|<0.819$~\cite{LumiPaper}. Assuming Chiral Belle achieves it's goal of a 70\% polarized $e^-$ beam, and taking \salr=0.00012 from ReneSANCe for the angular acceptance of $|\cos\theta|<0.819$, a measured \salr of $A^{\textrm{meas}}_{LR\Sigma}=\langle P_e\rangle A_{LR\Sigma}=0.000098$ is predicted, with a statistical uncertainty of 2.3\% for 40~ab$^{-1}$ of data. Dominant systematic uncertainties are expected to arise from knowledge of the beam polarization, background modeling, angular acceptance, and knowledge of the center-of-mass energy of the collisions. The \alr scales linearly with the beam polarization and uncertainty in it's measurement translates directly to measuring \salr. The \babar collaboration has demonstrated a technique to extract the beam polarization at the interaction point from tau decay kinematics with a systemic uncertainty of $\pm 0.0029$~\cite{babarpol}. This represents a relative uncertainty of 0.4\% for $\langle P\rangle=0.70$. 
The dominant backgrounds for this selection are expected to be a mix of $u\bar{u}, \tau^+\tau^-,$ and $d\bar{d}$ events which contribute
0.07\% of the selection~\cite{LumiPaper}. The $d\bar{d}$ events contribute the largest effect as they carry an asymmetry roughly two orders of magnitude times largely than the Bhabhas, $\salr^{d\bar{d}}\approx-0.020$. If we conservatively assume the entire background is composed of $d\bar{d}$, and the backgrounds are controlled to the 10\% level, a systematic uncertainty of 1\% is projected if the backgrounds are not further controlled.
The Belle~II calorimeter location is known to within 0.11~mm along the beam axis~\cite{Adachi_2025}. This corresponds to an acceptance uncertainty of 36 micro-radians or $2\times10^{-3}$ degrees. Propagating this uncertainty through ReneSANCe results in a 0.06\% uncertainty on \salr. The center-of-mass energy is expected to be known to within $\sim5$~MeV~\cite{Adachi_2025} and a similar evaluation in ReneSANCe results in a~0.07\% uncertainty in \salr. Adding all the sources of systematic uncertainty in quadrature gives a total systematic uncertainty estimate of 1\%, dominated by the background projections.

\subsection{Weak mixing angle}
\begin{figure}[]
    \centering
	\includegraphics[width=\linewidth]{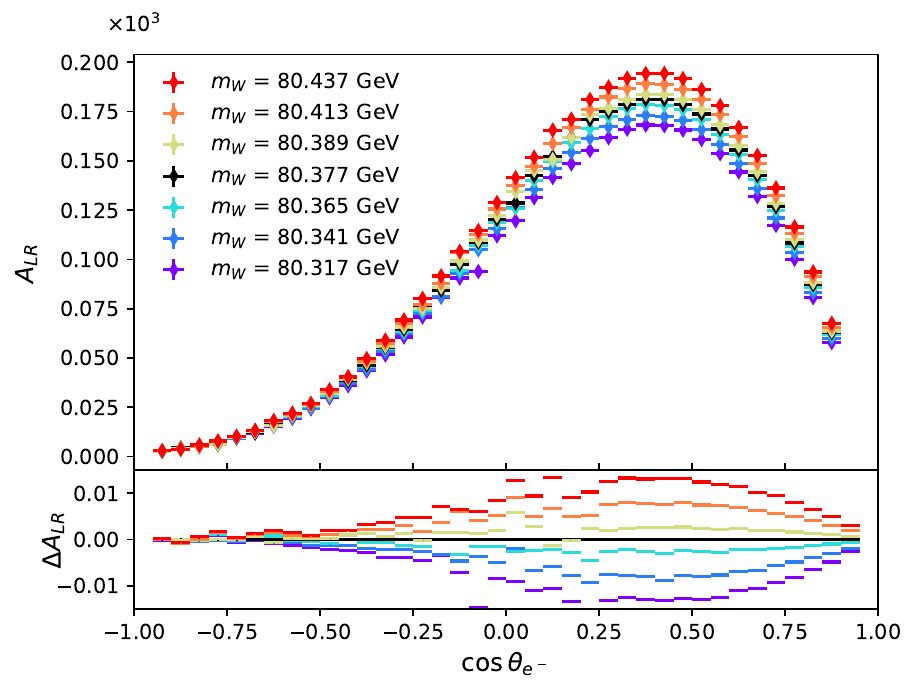}
	\caption{\alr in Bhabhas for bins of $\cos\theta=0.05$ across the angular acceptance in NLO. $m_W$ is input as 1$\sigma$, 3$\sigma$, and 5$\sigma$ deviations from the world average of $80.377\pm0.012$~GeV.}
	\label{fig:alrmw}
\end{figure}

\begin{figure}[]
    \centering
	\includegraphics[width=\linewidth]{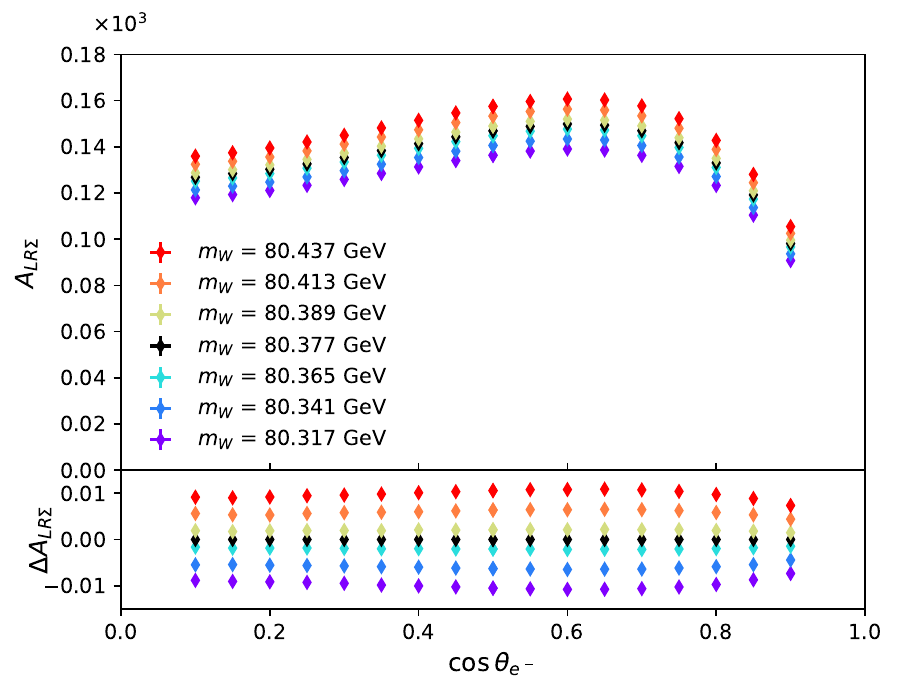}
	\caption{\salr in Bhabhas for an angular acceptance of $\theta<a<180^\circ-\theta$. $m_W$ is input as 1$\sigma$, 3$\sigma$, and 5$\sigma$ deviations from the world average of $80.377\pm0.012$~GeV.}
	\label{fig:swSense}
\end{figure}

In order to determine Chiral Belle's sensitivity to $\sin^2\theta_W$, the value of the mass of the W-boson ($m_W$) is varied in the generator, as recommended by the ReneSANCe authors~\cite{rene_private}. This correspondingly changes the value of $\sin^2\theta_W$, which can be  expressed at the Born level as $\sin^2\theta_W$=$1-\frac{m_W^2}{m_Z^2}$. As a baseline, the generator setup is changed to use the PDG value of $m_W$, 80.377 GeV~\cite{pdg}, and from the world average uncertainty of 12~MeV, $\pm1,3,5\sigma$ changes in the mass are applied to study the effect on \alr and \salr. The effects are shown for \alr in Figure \ref{fig:alrmw} and for \salr in Figure \ref{fig:swSense}


The expected sensitivity to \stw obtained by measuring \alr in each $\cos\theta$ bin is calculated from the change in the average \alr in each bin, $\Delta\alr$, associated with a change in \stw, $\Delta\stw$, in that bin as determined by varying the value of $m_W$. The overall selection efficiency of 0.3593 is assumed to apply to each bin. For each  
 $\cos\theta$ bin the projected uncertainty is $\sigma_{\stw}=\sigma_{\alr}\times\left(\Delta\stw /\Delta\alr\right)$. The value of $\sigma_{\alr}$ in each bin is determined from the cross-section in that $\cos\theta$ bin as calculated by ReneSANCe.  The overall \stw uncertainty is determined from the weighted mean of the values in all bins under the assumption of independent measurements. From these studies the weighted average value of $\left(\Delta\stw /\Delta\alr\right)$ is found to be 230. For 40~ab$^{-1}$, a 70\% polarization, and detector acceptance of $|\cos\theta|<0.90$,  the projected precision from the Bhabha \alr measurement is $\sigma_{\stw}=\pm0.00028$. 
 For $|\cos\theta|<0.819$,  the Belle II fiducial acceptance used for the luminosities studies of \cite{LumiPaper}, the uncertainty increases to $\sigma_{\stw}=\pm0.00032$.

This Chiral Belle uncertainty using only Bhabha events, and consequently only sensitive to the $Z^0$-electron coupling, is similar to the $\pm0.00026$ precision  of $\sin^2\theta_W$  determined at the $Z^0$ pole by the SLD collaboration from $A_{LR}$ measurements~\cite{ALEPH:2005ab} for the combined lepton states, which are dominated by the measurements sensitive  to the $Z^0$-electron couplings. Combining this SLD result with  the forward-backward Z-pole asymmetry measurements from LEP involving only electrons, the  $Z^0$-electron effective vector coupling constant was determined to be 
$g_{Ve}=-0.03816\pm 0.00047$ (see Table~7.7 in ~\cite{ALEPH:2005ab}),
corresponding to a combined SLD-LEP uncertainty of $\pm 0.00024$ on \stw at the Z$^0$ pole involving only the $Z^0$-electron coupling. 
We note that CMS recently reported preliminary results on the
forward-backward asymmetry in $e^+e^-$ events near the $Z^0$ pole that yield a precision of $\pm0.00041$ on \stw \cite{Khukhunaishvili:2024zxm}.
The projected Chiral Belle sensitivity with 40~ab$^{-1}$  is also comparable to the \stw projected uncertainty of the MOLLER experiment ($\pm 0.00028$) at the lower 100~MeV energy scale.

Note that Chiral Belle will also measure \stw using muons\cite{ALRmuons} and tau leptons as well as  b-quarks and c-quarks. Combined with the electron measurements discussed here, these different measurements will yield precision neutral-current universality tests, as well as an uncertainty on \stw assuming universality of $\pm0.00019$ taking into account the common uncertainty on $\langle P\rangle$.
 
\section{Conclusions}
The \salr studies on the Bhabha process show a negligible $4.4\times10^{-7}$ difference, corresponding to a relative difference of 0.3\%, between the theory work from Ref.~\cite{ALRee} and the ReneSANCe generator~\cite{renesance}. The data generated by ReneSANCe for this study are publicly available~\cite{miller_2025_15616559}. Projections for the Chiral Belle experiment suggest an \salr measurement with a $\pm2.3\%_{\textrm{stat}}\pm1\%_{\textrm{sys}}$ relative uncertainty is achievable with 40~ab$^{-1}$ of data. While the NLO effects are understood at a high level of precision, the effects of the NNLO contributions have yet to be calculated, and are expected to be significant in comparison to the experimental uncertainties.\\
\indent A study of the sensitivity to \stw indicates that with 40~ab$^{-1}$ of polarized data Chiral Belle would make a measurement only involving the $Z^0$-electron couplings with a statistics-dominated precision of $\pm0.00032$ or better. Combining this with Chiral Belle \salr measurements from muons and taus under the assumption of lepton universality would yield an overall uncertainty of $\pm0.00019$ on \stw. These precisions are similar to the sensitivities of the most precise previous experiments. As these past high precision measurements were performed at the $Z^0$ pole, the measurements enabled by Chiral Belle provide a unique probe for new physics through the running of  neutral current couplings.\\
\indent At this level of precision, combined with its measurements of the other fermion species, Chiral Belle will also provide world-leading measurements of the universality of the $Z^0$-fermion couplings. The projected level of precision motivates a full calculation of the NNLO contributions at 10.58~GeV in order to fully interpret future Chiral Belle measurements in the context of the SM. 

\section{Acknowledgments}
The authors would like to thank Renat Sadykov and Vitaly Yermolchyk for their efforts to extend the capabilities of the ReneSANCe generator to enable this study, as well as their insights into our analysis methods. The authors would further like to thank Aleksandrs Aleksevejs and Svetlana Barkanova for their insights and helpful comments. 

\bibliography{main}
\end{document}